\documentstyle[12pt]{article}
\begin{document}
\newcounter{sec}
\def\be{\begin{eqnarray}}
\def\ee{\end{eqnarray}}
\def\ba{\begin{array}}
\def\ea{\end{array}}
\def\pa{\partial}
\def\ga{\gamma}
\def\sr{\stackrel{\rightarrow}}
\def\sl{\stackrel{\leftarrow}}
\def\srp{\stackrel{\rightarrow}{\partial}}
\def\slp{stackrel{\rightarrow}{\partial}}
\renewcommand{\theequation}{\thesec.\arabic{equation}}
\setcounter{sec}{2}
\bigskip\bigskip\bigskip\bigskip
\medskip
\begin{center}
\section*{ Reconstruction of $SU(5)$ Grand Unified Model In Noncommutative 
Geometry Approach  }
%\baselineskip=0.0\baselineskip
Jianming Li\footnote{
Email address: lijm@itp.ac.cn} and  Xingchang Song 
\\
 Department of Physics, Peking University, Beijing 100871, China 
\end{center}

%(Received 31 January 1996)

%\vskip 2cm

\begin{abstract}

 Based on the  generalized gauge theory on $M^4\times Z_2\times 
Z_3$, we  reconstructed the realistic $SU(5)$ Grand Unified model by a suitable 
assignment of fermion fields.
  The action of   group elements $Z_2$  on fermion fields  is the charge 
conjugation while the action of $Z_3$ elements represent generation translation. 
 We find 
that to fit the spontaneous 
symmetry breaking and gauge hierarchy of $SU(5)$ 
model a linear term of curvature has to be introduced. A new mass relation is 
obtained in our reconstructed model. 
\end{abstract}

\vskip 1cm
{PACS number(s):  02.40. -k, 11.15. -q, 12.10. Dm} 
\newpage 
\section[toc_entry]{Introduction}

In recent years, 
it is believed that Non-commutative geometry extends the  basic 
geometry framework of physics\cite{connes,madore}. The most remarkable results 
are 
 that in Standard Model the 
Higgs fields  may be considered as a kind of gauge field by 
the same 
footing as Yang-Mills fields and the Yukawa couplings can be 
introduced as a kind of gauge coupling. These topics have been 
studied by many works\cite{conlot}---\cite{morita}.
It is also interesting to quest whether the same description stands when we go 
from Standard Model to Grand Unification theories (e.g. $SU(5)$ 
GUT\cite{georgi}), in which Higgs fields are introduced as input data in model 
building.
By enlarging the discrete points model first proposed by A.Connes\cite{connes, 
conlot}, A. Chamsedine et al \cite{ali}  provided a generalized formula, which 
gave a clue to study more extensive model beyond the 
Standard Model, such as $SU(5)$  and $SO(10)$ Grand unified 
models. But there are lots of details need to be further studied . 

In our previous works\cite{mine1,mine2}, we constructed generalized gauge theory 
on 
discrete group $Z_2$. In this approach,  we enlarged space-time to five 
dimensions with the 5-th ``coordinate" containing only two points of $Z_2$, 
assigned
left and right handed Fermion fields according to the  
discrete group 
``coordinate'' and wrote down a Lagrangian of  fermion 
fields, 
which is not only the function of the space--time coordinates but also of the  
discrete 
group "coordinate". The most 
important point of this approach was that the derivatives on discrete group were 
included in 
the Lagrangian.  Similar to   the case of the ordinary Yang--Mills gauge theory, 
when we require 
the 
Lagrangian be 
invariant under the gauge group which is a function of space time 
and of discrete 
group, then Higgs appear in the covariant derivative and 
Yukawa coupling is 
naturally introduced by the gauge coupling. Furthermore,we constructed the 
Weinberg-Salam model and  the Eleactroweak-strong interaction model and  
tried to 
endow discrete group with some physical meaning .  

In this paper, we first develop our previous approach to the case of $M^4\times 
Z_2\times Z_3$ 
and reconstruct the realistic $SU(5)$ 
Grand Unified model of three 
generation fermions with  generalized gauge theory on 
$M^4\times Z_2\times Z_3$. A similar generalized gauge theory on $M^4\times 
Z_2\times Z_3$   has also been discussed in a $CP$ violation toy 
model\cite{ding}. 
 We  distinguish  the left and right hand parts of fermions by 
two 
elements of the discrete group $Z_2$, differentiate three families by three 
elements of the discrete group $Z_3$ and connect fermions  by  charge 
conjugation 
transformation on discrete points of $Z_2$ and by generation translation on 
discrete points of $Z_3$.    Since there are   two mass scales in the $SU(5)$ 
model  
characterizing the spontaneous symmetry breaking of $SU(5)$ to $SU(3)\times 
SU(2)\times U(1)$ and then  to $SU(3)\times U(1)$, if we want to get this gauge 
hierarchy, we need to add the  linear term of  curvature $F$, which is first 
proposed by Sitarz\cite{sitarz}.

The plan of this paper is as follows. In section 2, we give the basic notion 
of 
gauge theory   on 
$M^4\times Z_2\times Z_3$. In section 3, we build $SU(5)$ model using the 
generalized gauge theory on $M^4\times Z_2\times Z_3$. In section 4, we discuss 
the 
symmetry broken phenomenon. 
 
\bigskip
\section[toc_entry]{Notation of gauge theory on $M^4\times Z_2\times Z_3$}
In this section we shall give the basic notion of construction 
gauge theory on 
$M^4\times Z_2\times Z_3$. More detailed account of 
construction can be found in 
\cite{sitarz, mine1}.

Let $x^\mu$ denote the coordinate on $M^4$ and $g$ label the 
points of discrete 
group  $Z_2\times Z_3$. The differentiation of an arbitrary function  on 
product space 
$M^4\times 
Z_2\times Z_3$ has the following form,
\be
 df=\partial_\mu f dx^\mu+\partial_g f \chi^g, g\in  
Z_2\times Z_3,\ee
where 
$dx^\mu$ and $\chi^g$ are basis of one forms on $M^4$ and  
$Z_2\times Z_3$ respectively.
The partial derivative $\partial_g$ is defined as follows:
\be
\partial_g f(x,h)=(f(x,h)-R_g f(x,h))=(f(x,h)- 
f(x,h\cdot g)).\ee
From the defination we can obtain a lot of ralations: ones for the product of 
one-forms 
\be\ba{cl}
&dx^\mu \hat\otimes dx^\nu=-dx^\nu \hat\otimes dx^\mu\\
&dx^\mu \hat\otimes \chi^g=-\chi^g\hat\otimes dx^\mu,\ea\ee
 ones for the multiplication of one-form by functions
\be
 f(x,h)dx^\mu =dx^\mu f(x,h),~~~\chi^g f(x,h)=R_g f(x,h) 
\chi^g,\ee
and ones by acting  derivative operator on one forms
$$ddx^\mu=0,~~d\chi^g=-C^g_{p,h} \chi^p\hat\otimes \chi^h,$$ 
where structure constants  
$C^g_{p,h}=\delta^g_p+\delta^g_h-\delta^g_{ph}(\delta^e_{ph}-1)$. 
The general gauge potential $A$ on $M^4\times Z_2\times Z_3$ 
can be written as:
\be
A=A_\mu dx^\mu+
\displaystyle\sum_{\tiny \ba{c}g\in Z_2\times Z_3\\ {g\neq 
e}\ea} \phi_g\chi^g\ee
The unitarity of gauge group enforces that $A^*=-A$. Thus, 
since 
$(dx^\mu)^*=dx^\mu$ and $({\chi}^{g})^{*}=-{\chi}^{g^{-1}}$, 
we obtain
$$(A_\mu)^{\dag}=-A_\mu,~~\phi_g^{\dag}=R_g \phi_{g^{-1}}.$$
The curvature two form 
$
F=dA+A\hat\otimes A$ splits into terms,
\be
F=\frac 1 2 F_{\mu\nu} dx^\mu\hat\otimes dx^\nu+F_{\mu 
g}dx^\mu\hat\otimes \chi^g+
F_{gh} \chi^g\hat\otimes \chi^h,\ee
where 
$$F_{\mu\nu}={\partial}_{\mu}A_{\nu}-{\partial}_{\nu}A_{\mu}+
[A_{\mu},A_{\nu}], $$
\be
F_{ \mu g}={\partial}_{\mu}\Phi_g+ A_{\mu}\Phi_g-\Phi_g 
R_{g}( A_{\mu}),
\ee
$$
 F_{gh}={\partial}_{g}{\phi}_{h}+{\phi}_{g}{R}_{g}{\phi}_{h}-
{{C}^k_{gh}}{\phi}_{k}\label{C} $$
with $\Phi_g=1-\phi_g$.

To construct the Yang-Mills action, we need to define  the metric 
\be\begin{array}{cl}
&<{dx}^{\mu},{dx}^{\nu}>=g^{\mu\nu},~~~~<{\chi}^{g},{\chi}^{h}
>=
{\eta}^{gh},\\[4mm]
&<{dx}^{\mu}\wedge {dx}^{\nu},{dx}^{\sigma}\wedge{dx}^{\rho}>=
\frac{1}{2}(g^{\mu\sigma}g^{\nu\rho}-g^{\mu\rho}g^{\nu\sigma})
,\\[4mm]
&<{dx}^{\mu}\otimes{\chi}^{g},{dx}^{\nu}\otimes{\chi}^{h}>=
g^{\mu\nu}{\eta}^{gh},\\[4mm]
 &<{\chi}^{g}\otimes {\chi}^{h},{\chi}^{g'}\otimes 
{\chi}^{h'}>=
{\eta}^{gg'}{\eta}^{hh'}.\end{array} \ee
where $\eta^{gh}=\eta_g \delta^{gh^{-1}}$.  After taking such a form
of the metric, the Yang-Mills Lagrangian becomes 
\be\ba{cl} {\cal
L}_N=&-{1\over N}\displaystyle\int_G <F,\overline{F}>\\ &={1\over
N}\displaystyle\int_G (-\frac 1 4 F_{\mu\nu}F^{\dag\mu\nu}+ \eta_g F_{\mu 
g}F^{\dag\mu}_{ g}-\eta_g
\eta_h F_{gh} F^{\dag}_{gh}) \ea\ee

 It was found\cite{sitarz} that there exist a
possibility of adding an extra gauge invariant term to the Yang-Mills action,
which is linear in the curvature $<F>$, 
\be \ba{cl} {\cal L}_{L}=&-{1\over N}\displaystyle\int_G<F>\\ &=-{1\over
N}\displaystyle\int_G F_{gh}\eta^{gh}= -{1\over N}\displaystyle\int_G \eta_g
F_{gg^{-1}}.\ea \ee 
Let us add this term to Yang-Mills action with an arbitrary
scaling parameter $\alpha$. We obtain the bosonic sector Lagrangian
\be {\cal
L}={\cal L}_N+\alpha {\cal L}_L.\ee
 In the next section, we will find that this 
Lagrangian 
is needed in the construction of the $SU(5)$ model.

\setcounter{sec}{3}
\setcounter{equation}{0}
\section[toc_entry]{Generalized $SU(5)$ gauge theory on $M^4\times Z_2\times 
Z_3$}
In this section, we build the  $SU(5)$ gauge theory  on $M^4\times Z_2\times 
Z_3$ by using generalized gauge theory on discrete group\cite{mine1, mine2}. 
According to the basic 
knowledge of $SU(5)$ model, we first set fermion fields on discrete group, then 
 write down gauge fields in terms of gauge potential, at last give Lagrangian of 
gauge 
fields by noncommutative  differential geometry approach.   
\subsection[toc_entry]{Fields on $M^4\times Z_2\times Z_3$}

From the basic knowledge of  $SU(5)$ model \cite{licheng}, we know that 
 one family of left(or right) handed  fermions can be accommodated 
in an $SU(5)$
 reducible representation of $5^*+10$(or $5+10^*$). 
According to representation of $SU(5)$, we write down the 
first family fermions 
as following:
 \be\ba{cl}
5^*:&\psi_L=\left[\ba{cl}&d_1^C\\&d_2^C\\&d_3^C\\&e^-\\
&-\nu_e\ea\right]_L,
~~~~~5:\psi^C_R=\left[\ba{cl}&d_1\\&d_2\\&d_3\\
&e^+\\&-\nu^C_e\ea \right]_R\\[15mm]
10:&\chi_L={1\over \sqrt{2}}\left 
[\ba{ccccc}0&u^C_3&-u^C_2&u_1&d_1\\
-u^C_3&0&u^C_1&u_2&d_2\\
u^C_2&-u^C_1&0&u_3&d_3\\
-u_1&-u_2&-u_3&0&e^+\\
-d_1&-d_2&-d_3&-e^+&0\ea\right]_L,\\[15mm]
10^*:&\chi^c_R={1\over \sqrt{2}}\left 
[\ba{ccccc}0&u_3&-u_2&u^C_1&d^C_1\\
-u_3&0&u_1&u^C_2&d^C_2\\
u_2&-u_1&0&u^C_3&d^C_3\\
-u^C_1&-u^C_2&-u^C_3&0&e^-\\
-d^C_1&-d^C_2&-d^C_3&-e^-&0\ea\right]_R.\ea\ee
The other two families can be written similarly  by replacing $u,d,e,\nu_e$ 
by $c,s,\mu,\nu_\mu$ and $t,b,\tau,\nu_\tau$.  From observation of  
 three 
family fermions and 
their left-right hand parts, $5^*+10$ and $5+10^*$, we find it 
is possible to 
assign them with respect 
to elements of discrete group $Z_2\times Z_3$, which is 
product of discrete 
groups $Z_2$ and $Z_3$. 
Discrete group $Z_2$ has two elements $Z_2=\{e,Z|Z^2=e\}$, 
discrete group $Z_3$ 
has three elements 
$Z_3=\{e,r,r^2|r^3=e\}$. So the direct product group $Z_2\times Z_3$ 
has six elements
$$Z_2\times Z_3=\{e,r,r^2Z,Z r,Z r^2|Z^2=e,r^3=e,Zr=rZ\}.$$
In this paper, we use two elements of $Z_2$ to distinguish 
left-right hand 
fermions and use three 
elements of $Z_3$ to distinguish three families. So our 
working manifold should be 
$M^4\times Z_2\times 
Z_3$.

According to discrete group $Z_2\times Z_3$,
we arrange Fermions as following:
\be\ba{cl}
&\psi(x,e)=\left[\ba{cl}\psi^C\\\chi^C\ea\right]^1_R,~~
\psi(x,r)=\left[\ba{cl}\psi^C\\\chi^C\ea\right]^2_R,~~
\psi(x,r^2)=\left[\ba{cl}\psi^C\\\chi^C\ea\right]^3_R,\\[5mm]
&\psi(x,Z)=\left[\ba{cl}\psi\\\chi\ea\right]^1_L,~~
\psi(x,rZ)=\left[\ba{cl}\psi\\\chi\ea\right]^2_L,~~
\psi(x,r^2 Z)=\left[\ba{cl}\psi\\\chi\ea\right]^3_L\ea,\ee
where $\left[\right]^i$ represents the $i$-th generation of 
Fermions.
It is important to note that the actions $R_g,g\in Z_2\times 
Z_3$ on 
fermions have definite physical meaning. We find that the 
action $R_Z$ is 
nothing but the charge 
conjugation transformation, which inter-changes left-right hand 
fermions in
$5^*+10$ and $5+10^*$ 
and the action $R_{r^i},i=1,2,3$ is the 
translation between 
different generations,
$$R_{r^i} \psi^j=\psi^{\left[(i+j)|Mod~ 3\right]},~~~~i=1,2; ~~j=1,2,3$$
As we did in \cite{mine1,mine2}, to build gauge theory on space $M^4\times 
Z_2\times Z_3$, we 
should introduce free fermion Lagrangian first,
\be\ba{cl}
{\cal L}(x,g)=&\overline{\psi}(g)\left [i\gamma^\mu
(\overrightarrow{\partial}_\mu-
{\overleftarrow\partial}_\mu)-U(\partial_Z+\partial_{Zr}+
\partial_{Zr^2})-U_1(\partial_{r}+\partial_{r^2})\right ]\psi(g),\\
&g\in 
Z_2\times Z_3\ea
\label {ffl}\ee
where  $U$, $U_1$ are parameters with mass dimension. Here we just choose two 
free parameters in front of partial derivatives of discrete group, a special 
case that 
depend on the model to be building, since these parameters are 
directly related to 
the mass 
of 
Higgs particles and  there are only two mass scales of Higgs fields in minimum 
$SU(5)$ 
model. In fact, $U$ and $U_1$ are parameters relate to the distance among 
discrete points in non-commutative geometry approach.
 
 Similar to the reason that leads to the introduction of 
Yang-Mills fields, it is 
reasonable to require that the Lagrangian (\ref {ffl}) be 
invariant under gauge 
transformations $H(x,g),g \in Z_2\times Z_3$, where $H$ are functions depending 
not 
only   
on $M^4$ but also on discrete group. So one should 
introduce covariant 
derivative in Lagrangian (\ref {ffl}).
 
Gauge invariant Lagrangian under $SU(5)$ group should be 
written as follows:
\be\ba{cl}
&{\cal L}_F(x,g)=\overline{\psi}(g)\left[i\gamma^\mu
(\overrightarrow{D}_\mu-
{\overleftarrow D}_\mu)-U(D_Z+D_{Zr}+
D_{Zr^2})-U_1(D_{r}+D_{r^2})\right]\psi(g),\\[5mm]
&D_\mu=\partial_\mu+ig A_\mu, ~~~~D_g=\partial_g+\phi_g R_g,~~g\in 
Z_2\times Z_3,
\label{fflc}
\ea
\ee
where 
\be
A(e)=\left[\ba{cc}\left( 
A_{k,l}\right)&\\&\left(A^*_{mn,pq}\right)\ea
\right],~~
A(Z)=\left[\ba{cc}\left( A^*_{k,l}\right)&\\&\left( 
A_{mn,pq}\right)\ea\right],
\ee
 $\left(A_{k,l}\right)$ is a $5\times 5$ matrix valued on $24$ generators of 
$SU(5)$ group and the 
corresponding
matrix elements are $A_{k,l}$;  $\left(A_{mn,pq}\right)$ is a
$25\times 25 $ 
matrix
 with $mn,pq$ denoting the row and column  indices of the matrix and the 
matrix element are 
 $$A_{mn,pq}=A_{m,p}\delta_{n,q}+A_{n,q} \delta_{m,p}.$$ 

Because the gauge transformations are independent of 
generations,  we should set 
Yang-Mills  potentials to be the same in different generations.This means 
$A(e)=A(r)=A(r^2)$ and $A(Z)=A(rZ)=A(r^2 Z)$.

In minimum $SU(5)$ model, there are two Higgs multiplets which belong to the  
adjoint and the
vector representations  respectively. Only the vector Higgs field 
appears in 
Yukawa coupling. In Yukawa terms of Lagrangian (\ref {fflc}), it is easy to find 
that $\phi_Z$, $\phi_{rZ}$, $\phi_{r^2 Z}$ connect left-right hand fermions and 
$\phi_r$,$\phi_{r^2}$ connect fermions with the same chirality. So only 
$\phi_Z$, 
$\phi_{rZ}$, $\phi_{r^2 Z}$ fields appear in Yukawa terms while 
$\phi_r$,$\phi_{r^2}$ fields do not. To get the minimum $SU(5)$ model, we 
arrange 
vector representation in $\phi_Z$, $\phi_{rZ}$, $\phi_{r^2 Z}$ and adjoint 
representation in  $\phi_r$, $\phi_{r^2}$. Thus  we write down the Higgs fields 
as 
following,
 $$\ba{ccc}
g=e&g=r&g=r^2\\[5mm]
\phi_Z(g)
=\tiny{\left[\ba{cc}0&f_{11}\left(H^*_{i,mn}\right)\\f_{11}
\left(H^*_{pq,j}\right)&e_{11}\left(H_{pq,mn}\right)\ea\right]};
&
\tiny{\left[\ba{cc}0&f_{22}\left(H^*_{i,mn}\right)\\f_{22}
\left(H^*_{pq,j}\right)&e_{22}\left(H_{pq,mn}\right)\ea\right]};
&
\tiny{\left[\ba{cc}0&f_{33}\left(H^*_{i,mn}\right)\\f_{33}
\left(H^*_{pq,j}\right)&e_{33}\left(H_{pq,mn}\right)\ea\right]}
\\[7mm]
\phi_{rZ}(g)
=\tiny{\left[\ba{cc}0&f_{21}\left(H^*_{i,mn}\right)\\f_{12}
\left(H^*_{pq,j}\right)&e_{12}\left(H_{pq,mn}\right)\ea\right]};
&
\tiny{\left [\ba{cc}0&f_{32}\left(H^*_{i,mn}\right)\\f_{23}
\left(H^*_{pq,j}\right)&e_{23}\left(H_{pq,mn}\right)\ea\right]};
&
\tiny{\left [\ba{cc}0&f_{13}\left(H^*_{i,mn}\right)\\f_{31}
\left(H^*_{pq,j}\right)&e_{31}\left(H_{pq,mn}\right)\ea\right]}
\\[7mm]
\phi_{r^2 Z}(g)
=\tiny{\left [\ba{cc}0&f_{31}\left(H^*_{i,mn}\right)\\f_{13}
\left(H^*_{pq,j}\right)&e_{13}\left(H_{pq,mn}\right)\ea\right]};
&
\tiny{\left [\ba{cc}0&f_{12}\left(H^*_{i,mn}\right)\\f_{21}
\left(H^*_{pq,j}\right)&e_{21}\left(H_{pq,mn}\right)\ea\right]};
&\tiny{
\left [\ba{cc}0&f_{23}\left(H^*_{i,mn}\right)\\f_{32}
\left(H^*_{pq,j}\right)&e_{32}\left(H_{pq,mn}\right)\ea\right]}
\ea$$ 
where $\left(H_{i,mn}\right)$ is a $5\times 25 $  matrix, 
$\left(H^*_{pq,j}\right)$ 
is a $25 \times 5$ matrix, $H_{pq,mn}$ is a $25 \times 25$ matrix 
and the elements 
are
$$H_{i,mn}=H_m\delta_{i,n}-H_n\delta_{i,m},$$
$$H_{pq,j}=H_p\delta_{q,j}-H_q\delta_{p,j},$$
$$H_{pq,mn}=\epsilon_{pqmnk} H_k$$

Higgs fields on discrete points $Z,rZ,r^2 Z$ may be defined by 
Hermitian 
condition
$\phi^{\dag}_g=R_g \phi_{g^{-1}}$, which are
$\phi^{\dag}_Z=R_Z \phi_Z, \phi^{\dag}_{rZ}=R_{rZ} \phi^{\dag}_{r^2 
Z}, 
\phi^{\dag}_{r^2 Z}=R_{r^2 Z}\phi_{rZ}$.

The other two components of Higgs fields $\phi_r, \phi_{r^2}$ 
are set as,
{ $$\ba{cccc}
&g=e&g=r&g=r^2\\[5mm]
\phi_r(g)=&I \tiny{\left[\ba{cc}t_1\left(\sum_{i,j}\right)&\\
&s_1\left(\sum^*_{pq,mn}\right)\ea\right]};
&I\tiny{\left[\ba{cc}t_2\left(\sum_{i,j}\right)&\\
&s_2\left(\sum^*_{pq,mn}\right)\ea\right]};
&I\tiny{\left[\ba{cc}t_3\left(\sum_{i,j}\right)&\\
&s_3\left(\sum^*_{pq,mn}\right)\ea\right]}
\ea$$
$$\ba{cccc}
&g=Z&g=rZ&g=r^2 Z\\[5mm]
\phi_r(g)=&I\tiny{\left[\ba{cc}t_1\left(\sum^*_{i,j}\right)&\\
&s_1\left(\sum_{pq,mn}\right)\ea\right]};
&I\tiny{\left[\ba{cc}t_2\left(\sum^*_{i,j}\right)&\\
&s_2\left(\sum_{pq,mn}\right)\ea\right]};
&I\tiny{\left[\ba{cc}t_3\left(\sum^*_{i,j}\right)&\\
&s_3\left(\sum_{pq,mn}\right)\ea\right]},
\ea$$
where $I=\sqrt {-1}$  and $t_i$  $s_i$ are real parameters,

$$\Sigma_{pq,mn}=\Sigma_{p,q}\delta_{q,n}+\Sigma_{q,n} 
\delta_{p,m}$$
$\left(\Sigma_{i,j}\right)$ is a $5\times 5$ traceless Hermitian matrix,i.e 
$\left(\Sigma_{i,j}\right)=
\left(\Sigma_{i,j}\right)^{\dag}$ and $Tr \sum$=0.
  
 The Hermitian condition $\phi^{\dag}_{r^2}=R_{r^2} \phi_r$ 
gives the
 values of $\phi_{r^2}$ on discrete points as,
$$\ba{cccc}
&g=e&g=r&g=r^2\\[5mm]
\phi_r(g)=&\tiny{-I\left[\ba{cc}t_3\left(\sum_{i,j}\right)&\\
&s_3\left(\sum^*_{pq,mn}\right)\ea\right]};
&\tiny{-I\left[\ba{cc}t_1\left(\sum_{i,j}\right)&\\
&s_1\left(\sum^*_{pq,mn}\right)\ea\right]};
&\tiny{-I\left[\ba{cc}t_2\left(\sum_{i,j}\right)&\\
&s_2\left(\sum^*_{pq,mn}\right)\ea\right]}
\ea$$
$$\ba{cccc}
&g=Z&g=rZ&g=r^2 Z\\[5mm]
\phi_r(g)=&-I\tiny{\left[\ba{cc}t_3\left(\sum^*_{i,j}\right)&\\
&s_3\left(\sum_{pq,mn}\right)\ea\right]};
&-I\tiny{\left[\ba{cc}t_1\left(\sum^*_{i,j}\right)&\\
&s_1\left(\sum_{pq,mn}\right)\ea\right]};
&-I\tiny{\left[\ba{cc}t_2\left(\sum^*_{i,j}\right)&\\
&s_2\left(\sum_{pq,mn}\right)\ea\right]}.
\ea$$

Actually, we impose a symmetry 
$$R_Z \phi=-\phi^{*}$$
in the assignments of the fields $\phi_r,\phi_{r^2}$. It is 
interesting to find 
that
this constraint corresponds to the discrete symmetry which was introduced in 
standard $SU(5)$ 
grand unification 
model.
In $SU(5)$ model, we  require the Higgs potential terms is 
invariant under 
transformation $H\rightarrow-H$, $\sum\rightarrow -\sum$, which 
can remove 
unwanted
terms in the potential.

\subsection[toc-entry]{Lagrangian of Model}

After taking the assignments of Yang-Mills fields and Higgs fields ,
now we are ready to write down the Lagrangian of fermionic 
sector form (\ref {fflc}) ,
 which include couplings
of gauge fields.
\be\ba{cl}
{\cal L}_F=&\displaystyle\sum_{A,k}\overline{\psi}_{k,A}i
\gamma^\mu{D}_\mu\psi_{k,A}+
\sum_{A,k,l}\overline{\chi}_{kl,A}i\gamma^\mu{D}_\mu\chi_{kl,A
}\\[5mm]
&\displaystyle+2\sum_{A,B}\sum_{pqklm}M_{1A,B}\overline{\chi^C}_
{pq,A}\chi_{kl,B
}
\epsilon_{pqklm}H_m+h.c.\\[5mm]
&\displaystyle+\sum_{A,B,k,l}M_{2A,B}\overline{\chi^C}_{kl,A}\psi
_{k,B}
H^{\dag}_l+h.c.\ea
\ee
where $A,B$ are generation indices, the other indices are that of $SU(5)$ group 
and $M_{1A,B}$ $M_{2A,B}$ are elements of matrix $$M_1=\left 
[\ba{ccc}f_{11}&f_{12}&f_{13}\\
f_{21}&f_{22}&f_{23}\\f_{31}&f_{32}&f_{33}\ea\right]$$
$$M_2=\left 
[\ba{ccc}e_{11}&e_{12}&e_{13}\\e_{21}&e_{22}&e_{23}
\\e_{31}&e_{32}&e_{33}\ea\right].$$

It is easy to show that 
\be
\overline{\chi^C}_{ij,A}\chi_{kl,B}\epsilon_{ijklm}H_m=
\overline{\chi^C}_{kl,B}\chi_{ij,A}\epsilon_{ijklm}H_m,\ee
so we  set $M_2$ to be symmetric matrix, i.e $e_{AB}=e_{BA}$, or 
$$M_2=\left 
[\ba{ccc}e_{11}&e_{12}&e_{13}\\e_{12}&e_{22}&e_{23}
\\e_{13}&e_{23}&e_{33}\ea\right].$$

The Lagrangian of bosonic sector may be derived from  the 
generalized differential 
calculation on $M^4\times Z_2\times Z_3$. Using those assignments of fields on 
discrete groups and the basic knowledge of non-commutative geometry, we can 
obtain the Lagrangian of gauge fields. In the calculation, for simplicity, we 
set $\eta_Z=\eta_{rZ}=G$ and  note $\eta_{r}=G_1$. Because the calculation is 
fairly 
cumbersome,  we only write the result here. 

\be\ba{clll}
{\cal L}_G &=&-{1\over N}<F,\overline{F}>\\[4mm]
&=& -{g^2\over 4N} 66 F_{\mu\nu} F^{\mu\nu}+
{16\beta\over N} {G\over U^2}D_\mu H^{\dag} D^\mu H\\[4mm]
&&+{4 \alpha\over N}{G_1\over U^2_1} Tr(D_\mu \Sigma^{\dag} D^\mu 
\Sigma)-
[V(H,\Sigma)+V(\Sigma)+V(H)]\ea\ee
where 
 $$\alpha=t^2_1+t^2_2+t^2_3+10(s^2_1+s^2_2+s^2_3),$$
$$\beta=Tr(2M_1M^{\dag}_1+3M_2M^{\dag}_2),$$
and 
$$\ba{cl}
D_\mu H&=(\partial_\mu+ igA_\mu )H\\
 D_\mu \sum&=\partial_\mu \sum +ig(A_\mu
 \sum -\sum A_\mu),\ea$$ 
which show that Higgs fields $H$ and $\sum$ are vector and adjoint 
representations of $SU(5)$ group.
Here we wrote gauge bosons, Higgs fields $\sum$ and $H$ in their 
matrix forms\cite{licheng} as
 \be
A={1\over \sqrt{2}}\left [\ba{ccccc}
&&&X_{1}&Y_{1}\\
{[G}&-2B/&{{\sqrt{30}]^\alpha}_\beta}&X_{2}&Y_{2}\\
&&&X_{3}&Y_{3}\\
X^{\dag}_{1}&X^{\dag}_{2}&X^{\dag}_{3}&W^3/\sqrt{2}+3B/\sqrt{30}&W^{\dag}\\
Y^{\dag}_{1}&Y^{\dag}_{2}&Y^{\dag}_{3}&W^{-}&W^3/\sqrt{2}+3B/\sqrt{30}\ea\right]
\ee
\be
\sum=\left [\ba{ccccc}
&&&\Sigma_{X1}&\Sigma_{Y1}\\
{[\Sigma_8 ]^\alpha}_\beta&-&2\Sigma_0/\sqrt 
{30}&\Sigma_{X2}&\Sigma_{Y2}\\
&&&\Sigma_{X1}&\Sigma_{Y1}\\
\begin{picture}(0,0)(1,0)
\put(-10,0){\line(1,0){215}}\put(111,-50){\line(0,1){115}} 
\end{picture}\\
\Sigma^{\dag}_{X1}&\Sigma^{\dag}_{X2}&\Sigma^{\dag}_{X3}&&\\[-1.5mm]
&&&[\Sigma_3 ]^r_s&+3\Sigma_0/\sqrt 30\\[-1.5mm]
\Sigma^{\dag}_{X1}&\Sigma^{\dag}_{X2}&\Sigma^{\dag}_{X3}&&\ea\right],\ee
\be
H=\left [\ba{cl}
H_{t_1}\\H_{t_2}\\H_{t_3}\\H_{d_1}\\H_{d_2}\ea\right].\ee
Before giving the expression of potential, we normalize the coefficient of 
dynamics terms in above Lagrangian, so we can take values of  normalization 
constant $N$ and metrics $G,G_1$ as follows:

$$N=66 g^2=16 \beta{G\over U^2}=4\alpha{G_1\over U^2_1}$$
$$G={33\over 8} {1\over \beta} g^2U^2$$
$$G_1={33\over 2} {1\over \alpha} g^2 U^2_1.$$ Then the Lagrangian of gauge 
fields 
becomes:
 \be\ba{cl}
{\cal L}_G=&-{1\over 4}  F_{\mu\nu} F^{\mu\nu}+
D_\mu H^{\dag} D_\mu H
+ Tr(D_\mu \Sigma^{\dag} D^\mu 
\Sigma)\\[5mm]
&- [V(H,\Sigma)+V(\Sigma)+V(H)],\ea\ee
and the potential is given as following,
$$V(\Sigma)=-m^2_1 
Tr\Sigma^2+\lambda_1(Tr\Sigma^2)^2+\lambda_2
Tr\Sigma^4$$
$$V(H)=-m^2_2 H^{\dag}H+\lambda_3(H^{\dag}H)^2$$
$$ V(H,\Sigma)=\lambda_4 (Tr\Sigma^2) 
H^{\dag}H+\lambda_5 H^{\dag} 
\Sigma^2 H$$
where
\be\ba{cl}
&m^2_1=g^2U^2_1({33\over 2\alpha}-{99\over 32} 
{\alpha\over\beta^2}
{U^4\over U^4_1})\\
&m^2_2=g^2U^2({33\over 4\beta}-66{1\over \alpha}{U^2_1\over 
U^2})\ea\ee

$$\ba{cl}
&\lambda_1={99\over 2}{g^2}Tr(SS^{\dag})\\
&\lambda_2={33\over 4}{g^2}Tr(TT^{\dag}+10TrSS^{\dag})\\
&\lambda_3={33\over 4}{g^2\over \beta^2}Tr\{[Diag(M_1 
M^{\dag}_1)]^2+
[Diag(M^{\dag}_1 M_1)]^2+2[Diag(M_2 M^{\dag}_2)]^2+
2[Diag(M^{\dag}_2 M_2)]^2\}\\ 	
&\lambda_4={33\over 8}{g^2\over \beta}Tr[Diag(M^{\dag}_1 M) 
T+
Diag(M_1 M^{\dag}_1) S+4 Diag(M_2 M^{\dag}_2) S]\\
&\lambda_5={33\over 8}{g^2\over \beta}Tr[Diag(M_1 
M^{\dag}_1 )S-2 
Diag(M_2 M^{\dag}_2) S-Diag(M^{\dag}_1 M) T].\ea$$
In the above expressions, we used the following notations,
$$T={1\over 
\alpha}\left [\ba{ccc}t^2_1
+t^2_3&&\\&t^2_1+t^2_2\\&&t^2_2+t^2_3\ea\right]
$$

$$S={1\over 
\alpha}\left 
[\ba{ccc}s^2_1+s^2_3&&\\&s^2_1+s^2_2\\&&s^2_2+s^2_3\ea\right].
$$
It is easy to show that $$Tr(T+10 S)=2;$$

For a $3\times 3$ matrix $M$, we define $Diag(M)$ as the diagonal 
part of $M$
$$Diag(M)=\left 
[\ba{ccc}M_{11}&&\\&M_{22}&\\&&M_{33}\ea\right].$$

To express above formulas in  a simple form, we redefine parameters by absorbing 
 some  constants in free parameter $U$ and $U_1$,
$$\mu={33 g^2 U\over \beta},~~~~~\mu_1={33 g^2 U_1\over \alpha},$$ 
and 
$$\ba{cl}
&\hat{s}_1={s^2_1+s^2_3\over \alpha},~~~~\hat{t}_1={t^2_1+
t^2_3\over \alpha},\\
&\hat{s}_2={s^2_1+s^2_2\over \alpha},~~~~\hat{t}_2={t^2_1+
t^2_2\over \alpha}\\
&\hat{s}_3={s^2_2+s^2_3\over \alpha},~~~~\hat{t}_3={t^2_2+
t^2_3\over \alpha}\ea,$$ where $\hat{s}_i,\hat{t}_i$ ($i=1,2,3$), 
is positive real
numbers. 

So we can write  $T$ , $S$ and $m^2_1$ and $m^2_2$ in  a simple form as,
$$T=\left 
[\ba{ccc}\hat{t}_1&&\\&\hat{t}_2\\&&\hat{t}_3\ea\right],~~~~S=\left 
[\ba{ccc}\hat{s}_1&&\\&\hat{s}_2\\&&\hat{s}_3\ea\right]$$

$$m^2_1=\mu^2_1({1\over 2}-{3\over 32} {\mu^4\over 
\mu^4_1}),~~~~m^2_2=\mu^2({1\over 4}-2{\mu^2_1\over \mu^2}).$$

In the previous calculation, we only take into account the term 
$<F,\overline F>$. It can be shown that in this case the Higgs potential can't 
give correct symmetry breaking 
mechanism of 
$SU(5)$ group. Fortunately, 
if the  term $<F>$ is 
introduced in the Lagrangian, we can get correct results.
It is easy to show that,
$$<F>= {16\over U^2} G\beta H^{\dag} H+4\alpha{G_1\over 
U^2_1}Tr(\Sigma^2)$$
We should introduce Lagrangian as follows
$${\cal L}=-{1\over N}(<F,\overline{F}>+q^{\prime}<F>).$$
 One finds that 
$$-{q^\prime\over N}<F>=-q^{\prime} H^{\dag} H-q^{\prime} 
Tr(\Sigma^2).$$

We set  $q^{\prime}= q^2  \mu^2_1$ and recalculate the 
Lagrangian 
of gauge fields and find that only coefficients $m^2_1,m^2_2$ are modified as,
\be\ba{cl}
m^2_1&=\mu^2_1({1\over 2}+q-{3\over 32} {\mu^4\over 
\mu^4_1})\\[5mm]
m^2_2&=\mu^2[{1\over 4}+(q-2){\mu^2_1\over \mu^2}].\ea\ee

\setcounter{sec}{4}\setcounter{equation}{0}
\section[toc-entry]{The Realistic $SU(5)$ Model and Higgs Mechanism}

In the last section, we have  completed the model building of  generalized gauge 
theory on $M^4\times Z_2 
\times Z_3$, where the  potential of Higgs fields is  derived directly from the 
calculation of non-commutative geometry. But some of crucial points need to be 
studied further, such as, does the potential  provide the desired  mechanism of 
gauge 
symmetries breaking, (i.e $SU(5)\longrightarrow SU(3)\times SU(2)\times   
U(1)\longrightarrow SU(3)\times U(1)$) and  if the results  suit to describe the  
physical phenomenon?  

 \subsection[toc-entry]{Realistic $SU(5)$ Model}
It is known that there are  two mass scales in $SU(5)$ model: masses of the 
$X$,$Y$ and 
of $W$ gauge boson 
masses. There exits a vast hierarchy of gauge symmetries, $M_X$ larger than 
$M_W$ by something like $12$ order of magnitude. In this section, we will show 
that   the model we built in last section may give rise the desired symmetries 
broken and gauge hierarchy, if  we impose the following conditions  among 
parameters,  
\be\ba{cl}
&\mu_1\ll\mu\\[5mm]
&F={30 \lambda_4+9\lambda_5\over 60\lambda_1+14\lambda_2} < 1\\[5mm]
&q={4+F\over 2(1-F)}.\ea \label{condition}\ee

 Now if the 
conditions (\ref {condition})
 is under  consideration,  we may write down the Bosonic part
  Lagrangian of the  
  model  as 

\be\ba{cl}
{\cal L}_G=
&-{1\over 4}F_{\mu\nu} F^{\mu\nu}+ 
D_\mu H^{\dag} D_\mu H
+ Tr(D_\mu \Sigma^{\dag} D^\mu\Sigma)\\[4mm]
 &+m^2_1 
Tr\Sigma^2-\lambda_1(Tr\Sigma^2)^2-\lambda_2
Tr\Sigma^4\\[4mm]
&+m^2_2 H^{\dag}H-\lambda_3(H^{\dag}H)^2-
\lambda_4 Tr\Sigma^2 
H^{\dag}H-\lambda_5 H^{\dag} 
\Sigma^2, 
\ea\ee
where
 $$m^2_1={5\over 2(1-F)}\mu^2_1,~~~m^2_2={1\over 4}\mu^2+F m^2_1,$$

\be\ba{cl}
\lambda_1=&{99\over 2}{g^2}Tr(SS^{\dag})\\
\lambda_2=&{33\over 4}{g^2}Tr(TT^{\dag}+10TrSS^{\dag})\\
\lambda_3=&{33\over 4}{g^2\over \beta^2}Tr\{[Diag(M_1 
M^{\dag}_1)]^2+
[Diag(M^{\dag}_1 M_1)]^2+2[Diag(M_2 M^{\dag}_2)]^2\\
&+
2[Diag(M^{\dag}_2 M_2)]^2\}\\
\lambda_4=&{33\over 8}{g^2\over \beta}Tr[Diag(M^{\dag}_1 M_1) 
T+
Diag(M_1 M^{\dag}_1) S+4 Diag(M_2 M^{\dag}_2)S]\\
\lambda_5=&{33\over 8}{g^2\over \beta}Tr[Diag(M_1 
M^{\dag}_1) S-2 
Diag(M_2 M^{\dag}_2) S-Diag(M^{\dag}_1 M) T]
,
\ea\label{lambda}
\ee
In those expressions, matrices $M_1$, $M_2$, $T$ and $S$ are defined as:

$$M_1=\left [\ba{ccc}f_{11}&f_{12}&f_{13}\\
f_{21}&f_{22}&f_{23}\\f_{31}&f_{32}&f_{33}\ea\right],~~~M_2=\left 
[\ba{ccc}e_{11}&e_{12}&e_{13}\\e_{21}&e_{22}&e_{23}
\\e_{31}&e_{32}&e_{33}\ea\right]$$

$$T=\left 
[\ba{ccc}\hat{t}_1&&\\&\hat{t}_2\\&&\hat{t}_3\ea\right],~~~S=\left 
[\ba{ccc}\hat{s}_1&&\\&\hat{s}_2\\&&\hat{s}_3\ea\right],$$
where $\hat{s}_i, \hat{t}_i$ are positive real numbers and 
 satisfy the  condition $\frac 1 2 Tr(T+10S)=1$ .

So far we have constructed a realistic $SU(5)$ model. Our 
next task is to research  whether it gives us the desired physical results.

\subsection[toc-entry]{Symmetry Breaking} 

Since for parameters $\lambda_1$ and $\lambda_2$ in (\ref 
{lambda}), the 
following conditions are true, 
$\lambda_2>0$ and  $\lambda_1>-{7\over 30} \lambda_2$, 
potential $V(\Sigma)$ 
reachs its  minimum at
$$
\Sigma_0=V_1\left [\ba{ccccc}
2&&&&\\&2&&&\\&&2&&\\&&&-3&\\&&&&-3\ea\right]$$
where
$V^2_1={m^2_1\over 60 \lambda_1+14 \lambda_2}$, which was  derived by 
Li\cite{li}.
For the first stage, $SU(5)$ gauge symmetry  is spontaneous broken down to  
$SU(3)\times SU(2)\times U(1)$ as the scalar $\Sigma$ develops VEV, 
$<\Sigma>=\Sigma_0$. Because $\Sigma$ is a scalar in the adjoint representation 
of 
$SU(5)$,  mass terms for the $G^\alpha_\beta$, $W_r$, B fields remain to be 
zero, 
while  
the $X$ and $Y$ bosons acquire their  masses 
$$M_X=M_Y=\sqrt{{25\over 2}}g V_1$$.

For the second stage, gauge symmetries $SU(3)\times SU(2)\times U(1)$
 are broken 
to $SU(3)\times U(1)$ as scalar field $H$ takes its VEV as
   $$
<H>={1\over 2}\left [\ba{cl}0\\0\\0\\0\\V_2\ea\right],$$
where
$V^2_2={m^2_d\over \lambda_3}$,

$$m^2_d=m^2_2-(30 \lambda_4+9 \lambda_5) V^2_1={1\over 4} \mu^2.$$ 
Then bosons $W$ and $B$ obtain masses,
$$M_W={1\over 2} g V_2, ~~~M_B= \sqrt{2\over 5}g V_2.$$

Meanwhile Higgs fields also obtain their masses in this model, their values are 
listed in 
the 
following table,
 \be\ba{cc}
\begin{picture}(0,0)(1,0)
\put(-40,0){\line(1,0){160}} \end{picture}\\
Scalar~~ fields&[mass]^2\\
\begin{picture}(0,0)(1,0)
\put(-40,0){\line(1,0){160}} \end{picture}\\
{[\Sigma_8]^\alpha}_\beta&20\lambda_2V^2_1\\
{[\Sigma_3]^\alpha}_\beta&80\lambda_2V^2_1\\
\Sigma_0&4m^2_1\\
H_{t_\alpha}&\lambda_3 V^2_2+5 \lambda_5 V^2_1\\
H_{d_r}&\lambda_3 V^2_2\\
\begin{picture}(0,0)(1,0)
\put(-40,0){\line(1,0){160}} \end{picture}\ea\ee

It is interesting to note that
 $${m^2_d\over m^2_W}={33\over \beta^2} Tr \{ [Diag(M_1 
M^{\dag}_1)]^2+
 [Diag(M^{\dag}_1 M_1)]^2+2[Diag(M_2 M^{\dag}_2)]^2+
 2[Diag(M^{\dag}_2 M_2)]^2\}$$
 is a quantity which depends on fermionic mass matrix. This relation does not  
exist in the original $SU(5)$ Grand Unified Model.

Because parameters $\mu$ and $\mu_1$ were  chosen to be $\mu\ll\mu_1$ in 
conditions (\ref {condition}), it is easy to find $V_2\ll V_1$ in VEV, 
which means that we may realize   the masses of gauge bosons $X$ and $Y$  to be 
as heavy as 12 order of that of gauge bosons $W$ and $B$. Therefore the gauge 
hierarchy problem is fitting here. 
In fact, to realize $\mu\ll\mu_1$, we should take $U\ll U_1$ in the fermion 
lagrangian (\ref {fflc}).  From the point view of non-commutative geometry 
approach,  $U$ 
is a paramter labeling the distant between two discrete points of $Z_2$ and 
$U_1$ is that  labeling the distant among three discrete points of $Z_3$, 
These two geometry quantities control the mass scales of symmetries broken in 
our model.

\section[toc-entry]{Concluding remarks}
We have first constructed a  $SU(5)$   model with generalized gauge theory on 
$M^4\times Z_2\times Z_3$. We have shown that
the Higgs
mechanism is automatically included in the generalized gauge theory by 
introducing the Higgs fields as a kind of gauge fields with respect to the 
discrete groups and the Yukawa 
couplings   automatically given by the generalized 
gauge 
coupling principle.   
Then we arrange the parameters  appropriately and obtain the minimum $SU(5)$ 
grand unified model.
In the model, the  Higgs potential  can lead to the spontaneous symmetry broken 
mechanism of  
$SU(5)\longrightarrow  SU(3)\times SU(2)\times U(1)\longrightarrow SU(3)\times 
U(1)$, and they took place  in two different gauge hierarchy scalars. There  are 
also two scalars $H$ and $\sum$,  the vector and adjoint representations of 
$SU(5)$ group to break down gauge symmetry and enable the particles massive. In 
construction of the model, we arrange $H$ and $\sum$ in the connection matrices 
$\phi_Z$, $\phi_{Zr}$, $\phi_{Zr^2}$, $\phi_r$ and $\phi_{r^2}$. I want to 
emphasize that this assigements is unique in general, if they are set in a 
``wrong'' place, their transformation properties under $SU(5)$ group will not be 
satisfied.  It is 
worthy while to point out that the hierarchy scalars  depends on two geometry 
quantities, i.e.the distant of two discrete points in $Z_2$ and that of three 
discrete points in $Z_3$. One of the interesting starting point of this approach 
is to understand the  discrete groups $Z_2$ and $Z_3$ as charge conjugation 
transformation and generation translation in the free fermion lagrangian, 
although they are broken after the arrangements of  gauge fields. This is 
completely different from preious work.  

 There exist some differences  between the parameters of  the reconstructed 
model and the standard $SU(5)$ grand unified model. In Standard $SU(5)$ Model, 
there are following free parameters:
 \begin{itemize}

\item $g$--- SU(5) coupling constant,

\item $M_1,M_2$--- mass matrices, 

\item $m^2_1,m^2_2,\lambda_1,\lambda_2,\lambda_3,\lambda_4,\lambda_5$--- 
parameters in potential.

\end{itemize} 
In our reconstructed model, coupling constant $g$, mass matrices $M_1, M_2$ , 
and 
$m_1,m_2$ are also free parameters, instead of parameters 
$\lambda_1,\lambda_2,\lambda_3,\lambda_4,\lambda_5$, we introduced two matrices 
$S$ and $T$ and they
 satisfy condition $\frac 1 2 Tr(T+10S)=1$. On apparent observation,  
the number of parameters is the same in these two models, but now parameters 
$\lambda_1,\lambda_2,\lambda_3,\lambda_4,\lambda_5$ are functions of 
$M_1,M_2,S,T$ in the reconstructed model, so they are not as free as in the 
standard 
$SU(5)$ model. One result 
of this property is that the ratio of $M_{H_d}/M_W$ is a function of mass 
matrices 
which means there exists a complex relation  among the masses of 
particles at tree level. 
Therefore, it 
needs 
to be studied further   whether there are more relations. This approach is also  
available to study more extensive models such as the left-right symmetry 
model, the 
$SO(10)$ grand unified model and the supersymmetry model etc. 
 We will study these issues elsewhere.
 \bigskip 
\bigskip 

\centerline{\Large \bf Acknowledgement}
\bigskip

\bigskip
This work is supported in part by the National Science Foundation and Chinese 
Post Doctoral Foundation.
The authors  would like to thank Prof. H-Y Guo, K. Wu and Z.Y. Zhao for helpful 
discussions 
and Dr. C. Liu for useful comments.

\end{document}